\documentclass{ws-ijmpe}

\begin{document}

\markboth{A. V. Afanasjev \it{et. al}}{Conference proceedings}

\catchline{}{}{}{}{}

\title{Recent progress in the study of fission barriers in covariant density functional
theory}

\author{\footnotesize A. V. Afanasjev\footnote{Based on talk presented
at 18th Nuclear Physics Workshop ``Maria and Pierre Curie'', 2011, Kazimierz,
Poland}, H. Abusara}

\address{Department of Physics and Astronomy, Mississippi State University, Mississippi 39762
USA
}
\author{\footnotesize P. Ring}

\address{Physik-Department der Technischen Universit\"at M\"unchen
D-85748 Garching, Germany}

\maketitle

\begin{history}
\received{(received date)}
\revised{(revised date)}
\end{history}

\begin{abstract}
Recent progress in the study of fission barriers of actinides and
superheavy nuclei within covariant density functional theory
is overviewed.
\end{abstract}

\section{Introduction}

  A study of the fission barrier heights $B_{f}^{st}$
of nuclei is motivated by the importance of this quantity for
several physical phenomena. For example, many heavy nuclei decay by
spontaneous fission, and the size of the fission barrier is a measure
for the stability of a nucleus reflected in the spontaneous fission
lifetimes of these nuclei\cite{SP.07}. The $r-$process of stellar
nucleosynthesis depends (among other quantities such as masses and
$\beta$-decay rates) on the fission barriers of very neutron-rich
nuclei\cite{AT.99,MPR.01}. In addition, the population and survival
of hyperdeformed states at high spin also depends on the fission
barriers\cite{DPS.04,AA.08}.

  The physics of fission barriers is also intimately connected with
on-going search for new superheavy elements which is motivated by the
attempts to provide the answers for two open questions in nuclear structure,
namely, the limits of the existence of atomic nuclei at large
values of proton number and the location of the island of stability of superheavy
nuclei and the next magic numbers (if any) beyond $Z=82$ and $N=126$. 
For example, the probability for the formation of a superheavy
nucleus in a heavy-ion-fusion reaction is directly connected to the height
of its fission barrier\cite{IOZ.02} which is a decisive quantity in the
competition between neutron evaporation and fission of a compound nucleus
in the process of its cooling. The large sensitivity of the cross section
$\sigma$ for the synthesis of the fissioning nuclei on the barrier height
$B_{f}^{st}$ also stresses a need for accurate calculations of this value.

\begin{figure}
\centerline{\psfig{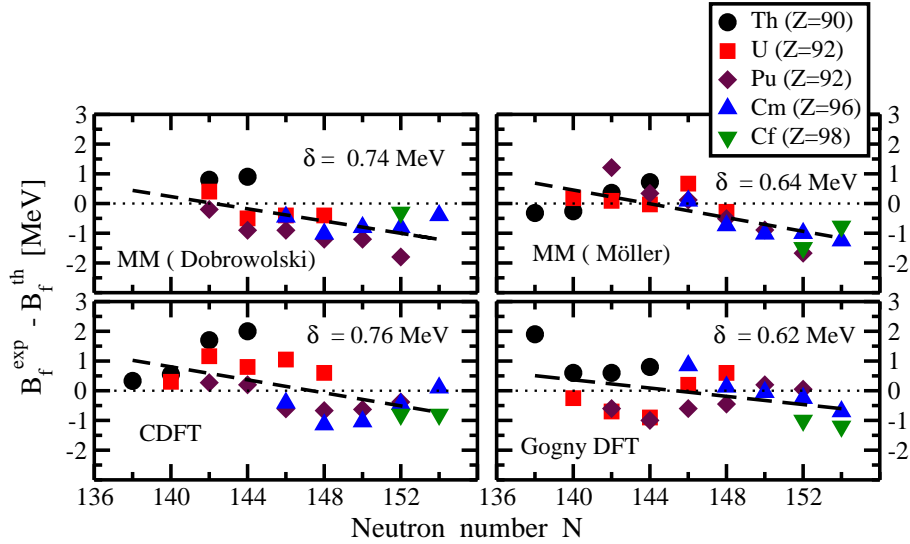}}
\vspace*{8pt}
\caption{The difference between experimental and calculated heights of
inner fission barriers as a function of neutron number $N$. The results of
the calculations are compared to estimated fission barrier
heights given in the RIPL-2 database \protect\cite{RIPL-2}, which is used
for this purpose in the absolute majority of theoretical studies on fission 
barriers in actinides. The results
of the calculations within microscopic+macroscopic method ('MM(Dobrowolski)'
\protect\cite{DPB.07} and 'MM(M{\"o}ller)' \protect\cite{MSI.09}), covariant
density functional theory ('CDFT' \protect\cite{AAR.10} and density functional
theory based on the finite range Gogny force ('Gogny DFT' 
\protect\cite{DGGL.06}) are shown. Thick dashed lines are used to show the 
average trend of the deviations between theory and experiment as a function 
of neutron number. The average deviation per barrier $\delta$ [in MeV] is 
defined 
as $\delta = \sum_{i=1}^N |B_f^i(th)-B_f^i(exp)|/N$, where $N$ is the number 
of the barriers with known experimental heights, and $B_f^i(th)$ 
($B_f^i(exp)$) are calculated (experimental) heights of the barriers. 
Long-dashed lines represent the trend of the deviations between theory and
experiment as a function of neutron number. They are obtained via linear
regression based on a least square fit.
}
\label{N-dep}
\end{figure}

  Covariant density functional theory (CDFT)\cite{VALR.05} provides
a natural incorporation of spin degrees of freedom and an accurate
description of spin-orbit splittings, which has an essential influence
on the underlying shell structure. Lorentz covariance of the CDFT
equations leads to the fact that time-odd mean fields are determined
with the same constants as time-even fields\cite{AA.10}. In addition,
pseudo-spin symmetry finds a natural explanation in the
relativistic framework\cite{Gin.97}. As a result, CDFT provides an
attractive framework for the description of the
structure of nuclei. Over the years a large variety of nuclear phenomena
have been successfully described within CDFT (see Ref.\cite{VALR.05}
and references therein). However, until recently most of the calculations
were restricted to axial symmetry and therefore it was not possible to provide
an accurate description of fission barriers. Only the inclusion of triaxiality in
Ref.\cite{AAR.10} has resolved this puzzle. This manuscript will review
recent progress in understanding the fission barriers in CDFT.

\section{Fission barriers in actinides}
\label{Fis-act}

\begin{figure}
\centerline{\psfig{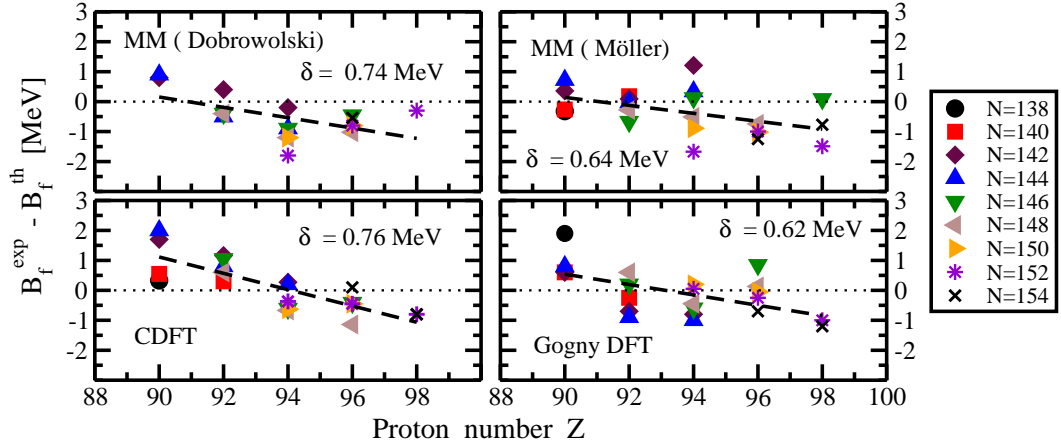}}
\vspace{-0.3cm}
\caption{The same as in Fig.\ \ref{N-dep} but
as a function of proton number $Z$.}
\label{Z-dep}
\end{figure}

  Earlier calculations performed within axially symmetric
relativistic mean field (RMF)+BCS\cite{BBM.04} and  relativistic
Hartree-Bogoliubov (RHB)\cite{KALR.10} frameworks have shown substantial
deviations from experiment for the heights of the inner fission barriers.
It is known from non-relativistic calculations that triaxiality
lowers the saddle in actinides and brings the results of calculations
closer to experiment (see discussion in Ref.\cite{AAR.10}). Although
a triaxial RHB computer code with finite range Gogny forces in the
pairing channel was available for more than 10 years\cite{ARK.00},
systematic calculations of fission barriers within this framework are
numerically too expensive. This is because a large basis including full
$N_F=20$ fermionic and $N_B=20$ bosonic shells as well as 24 mesh points
for the numerical Gauss-Hermite integration are needed for
the required accuracy\cite{AAR.10}. This problem
has been resolved in Ref.\cite{AAR.10} by performing calculations within
triaxial RMF+BCS model employing monopole pairing forces instead of
full RHB calculations. These simplifications decreased the computational 
time by at least of one order of magnitude, thus making systematic 
calculations of inner fission barriers feasible.

\begin{figure}
\centerline{\psfig{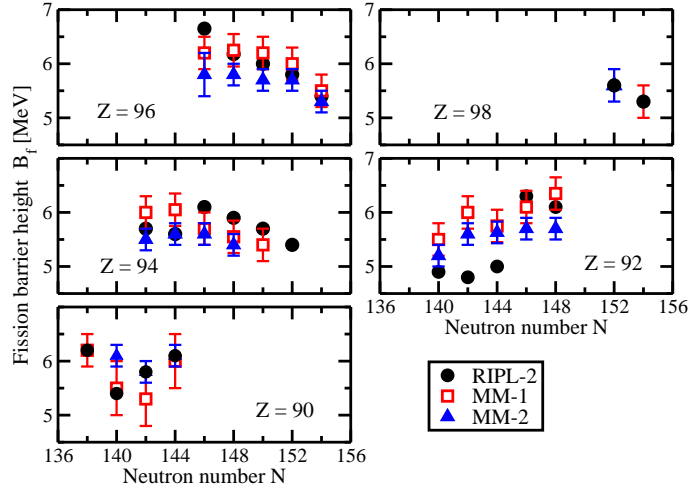}}
\vspace{-0.3cm}
\caption{The heights B$_f$ of inner fission barriers of actinides.
The results of the RIPL-2\protect\cite{RIPL-2}, 'MM-1' and
'MM-2'\protect\cite{MM.11} compilations are shown. Note that 'MM-1' is
based on Refs.\protect\cite{BHBG.74,B.80}, while 'MM-2' on
Ref.\protect\cite{BL.80}.}
\label{Exp-com}
\end{figure}

The inclusion of triaxiality has improved the accuracy of the
description of inner fission barriers in CDFT\cite{AAR.10} bringing
it up to the level of the accuracy typical for microscopic+macroscopic
(MM) approaches and DFT with Gogny forces. Since the results of this investigation
performed with the NL3* parametrization of the RMF Lagrangian\cite{NL3*}
are already published\cite{AAR.10} we will concentrate in this section
on the comparative analysis of the results obtained in different
theoretical frameworks.

   Figs.\ \ref{N-dep} and \ref{Z-dep} show the differences between
experimental and calculated heights of inner fission barriers obtained
in different theoretical models as a function of neutron and proton
numbers, respectively. Note that this comparison covers only results
of systematic triaxial calculations of even-even Th, U, Pu, Cm and Cf nuclei.  
To our knowledge, no such calculations have been published with DFT based
on Skyrme forces. As a result, these figures cover all existing systematic 
triaxial studies of inner fission barriers in actinides.

  The $\delta$-values displayed on the panels of Figs.\ \ref{N-dep} and
\ref{Z-dep} show the average deviation from experiment for the calculated 
heights of inner fission barriers. One can see that they are of the same magnitude 
in the different approaches and minor differences between the approaches in the 
$\delta$-values are not important considering the considerable uncertainties 
in the extraction of inner fission barrier heights from experimental data 
(see Fig.\ \ref{Exp-com}).

   However, the similarity of the average trends of these deviations (shown by thick
dashed lines in Figs.\ \ref{N-dep} and \ref{Z-dep}) as a function of
neutron and proton numbers is more important considering the differences in
underlying mean fields and in the treatment of pairing correlations.
At present, it is difficult to find a clear explanation for these trends.
Although differences in the treatment of pairing correlations (BCS
with monopole pairing and of different pairing windows in the
CDFT\cite{AAR.10} and MM\cite{DPB.07,MSI.09} calculations versus the
Hartree-Fock-Bogoliubov framework based on the D1S force in
Gogny DFT\cite{DGGL.06}) can contribute to deviations between theory
and experiment\cite{KALR.10}, it is quite unlikely that they are responsible
for the observed trends of the deviations. 

\section{The impact of the accuracy of the description of the single-particle
 energies on fission barriers}

An essential difference between the phenomenological models based on
Woods-Saxon or Nilsson potentials and self-consistient DFT calculations 
is the fact that the phenomenological potentials are fitted to experimental 
single-particle energies while, apart from the adjustment of the spin-orbit
force in the Gogny functional, no single particle information is used in the 
fit of the DFT parametrizations. As a consequence, the phenomenological potentials 
accurately describe the single-particle spectra
in deformed systems. On the contrary, several restricted in scope investigations
of experimental spectra in deformed odd nuclei\cite{A250,SDMMNSS.10,RSR.10}
showed that many of the DFT models do not possess the same spectroscopic
quality for the description of single-particle spectra achievable in the MM
method. A recent systematic investigation of the spectra of rare-earth
nuclei and actinides in CDFT\cite{AS.11} strongly enforces this conclusion.

\begin{figure}
\centerline{\psfig{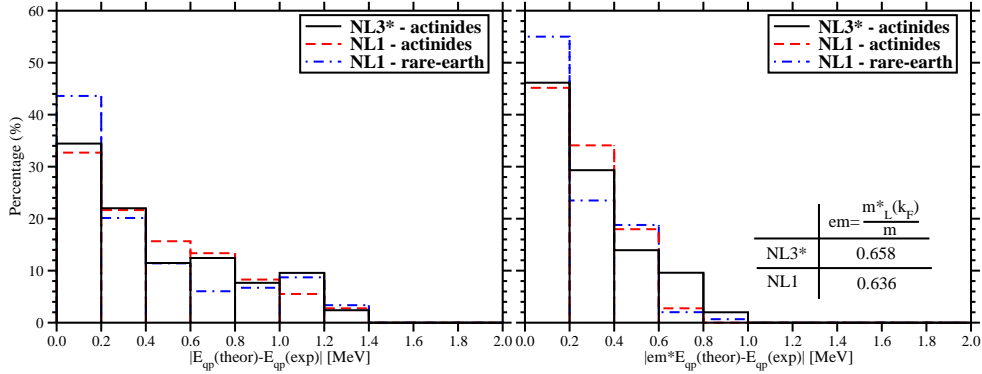}}
\vspace{-0.3cm}
\caption{(left panel) The distribution of the deviations of the calculated
energies $E_{qp}(theor)$ of one-quasiparticle states from experimental
ones $E_{qp}(exp)$. The vertical axis shows the percentage of the states
which deviate from experiment by the energy deviation range (the width of
bar) specified on horizontal axis. (right panel) The same as in left panel,
but for the case when the energy scale of theoretical spectra is corrected
for low Lorentz effective mass.}
\label{Dev-stat}
\end{figure}

  A statistical analysis\cite{AS.11} of discrepancies between calculated and
experimental energies of one-quasiparticle states in the ground state minimum
is presented in the left panel of Fig.\ \ref{Dev-stat}. One can see that in the
actinide region only approximately 33\% of one-quasiparticle states are
described with an accuracy better than 200 keV, and approximately 22\% with
an accuracy between 200 and 400 keV in the NL3* and NL1 parametrizations of the
RMF Lagrangian. The percentage of states for a given range of deviations
gradually decreases with increasing deviation between experiment and 
calculations. However, for some states the deviation of the calculated energy
from experiment exceeds 1 MeV and can be close to 1.4 MeV. Fig.\
\ref{Dev-stat} also shows that with the NL1 parametrization the 1-qp
energies in odd-proton rare-earth nuclei are somewhat better described as
compared with actinide region. Otherwise, the distribution histograms for
the deviations are similar in both regions and for both parametrizations.

  In the light of these results, it is important to understand why MM and
DFT models describe the experimental fission barriers with similar accuracy
(Sect.\ \ref{Fis-act}).

 {\it First}, it is important to remember that theoretical single-particle
energies can be confronted with experiment only at the ground state 
since reliable experimental data on the energies of dominant single-particle 
states either at the saddle or at the second (superdeformed) minimum are 
absent despite the fact that fission isomers in actinides have been 
observed almost 50 years ago\footnote{This is also
true for all regions of superdeformation studied so far in experiment.}.
Thus, both types of models (MM and DFT) rely in the study of fission
barriers on the ``extrapolations'' of the energies of single-particle
states to large deformations. Since the quality of such ``extrapolations'' is
not known one cannot say a priori that the description of the
single-particle energies in MM models is better than in DFT.

 {\it Second}, there exist important differences between the MM models and DFT in
the interpretation of the meaning of theoretical single-particle states
and their energies and neither of these interpretations takes fully into
account the complicated physics of single-particle degrees of freedom.

  It is well known that experimental ``single-particle''
states are not mean-field states. In reality, their wave functions are
fragmented and always contain the admixtures from vibrational 
phonons. In odd mass nuclei, the weights of these admixtures increase with
increasing excitation energy of the level relative to the ground
state\cite{Sol-book2}. By fitting
the parameters of phenomenological potentials
to the energies of dominant single-particle states, the MM models effectively
include vibrational corrections into these potentials but only on the
level of the energies and not on the level of the wavefunctions. As a
consequence, these potentials are characterized by an effective mass
of the nucleon at the Fermi level $m^*(k_F)/m\approx 1.0$ which reproduces
a calculated level density close to experiment.

  On the contrary, in density functional methods single-particle levels 
are not adjusted to experiment since their parameters are fitted mainly to 
bulk and neutron matter properties. As a consequence, most of them, in 
particular Gogny and relativistic functionals, are characterized by low 
effective
mass of the nucleon (the Lorentz mass for the case of CDFT), and calculated
single-particle states do not include vibrational corrections. A low
effective mass leads to a stretching of the theoretical single-particle
energy scale as compared with experiment, and, thus, to a larger
deviations between theory and experiment for deformed one-quasiparticle
states (left panel of Fig.\ \ref{Dev-stat}). To cure this problem
one should go beyond the mean field approximation and supplement CDFT by
particle-vibrational coupling (PVC)\cite{RW.73}. So far, this has been done 
only in spherical nuclei in Refs.\cite{LR.06,LA.11}, in which it was
shown that in the presence of PVC (i) calculated spectra of dominant
single-particle states come closer to experimental ones and (ii)
effective mass of the nucleon comes closer to 1.

  A similar compression of calculated spectra is expected also
in deformed nuclei. However, so far, no PVC model based on the DFT 
framework has been developed for such nuclei. The analysis of 
Ref.\cite{AS.11} suggests that on average the expected compression of
single-particle spectra can be achieved via a rescaling of 
one-quasiparticle (1-qp) energies by the Lorentz effective mass. The
impact of such an energy rescaling on the distribution of the deviations
between theory and experiment is shown in the right panel of Fig.\
\ref{Dev-stat}.  One can see that more than 75\% of states are described
with an accuracy better than 400 keV; this is a typical accuracy of the
description of the energies of deformed 1-qp states within
phenomenological potentials\cite{JSSJ.90,PS.04}. Although this energy
rescaling is somewhat schematic and assumes that the effect of
PVC is identical in spherical and deformed nuclei, it clearly 
indicates that PVC, leading to an increase of the effective mass, 
will also improve the description of experimental spectra as 
compared with mean field results.

\section{Extrapolation to superheavy nuclei}

   Systematic CDFT calculations of fission barriers allowing for triaxial
deformation are performed for even-even superheavy nuclei with charge
numbers $Z=112-120$ and $N-Z=48-62$ using three classes of models\cite{AAR.11}. 
These are nonlinear meson-nucleon coupling, the density-dependent meson-nucleon 
coupling and density-dependent point coupling models. The main differences between 
them lay in the treatment of the range of the interaction, the mesons and in the density
dependence. The interaction in the first two classes has a finite range,
while the third class uses zero-range interaction with an additional
gradient term in the scalar isoscalar channel. Mesons are absent in
the density-dependent point coupling model. It turns out that the
results obtained with these models are similar, so they will be
discussed on the example of the nonlinear meson-nucleon coupling model
represented by the NL3* parametrization.

 The calculations show that inner fission barriers of the Z=112 and 114 isotopes
are axially symmetric. In the Z=116 isotopes, the triaxiality has an
impact on inner fission barriers of nuclei with $N\geq 178$.  The role of
triaxiality becomes even more pronounced in the Z=118 and 120 isotopes,
especially the ones with neutron number close to $N=184$.

\begin{figure*}[ht]
\begin{center}
\includegraphics[width=6.3cm,angle=0]{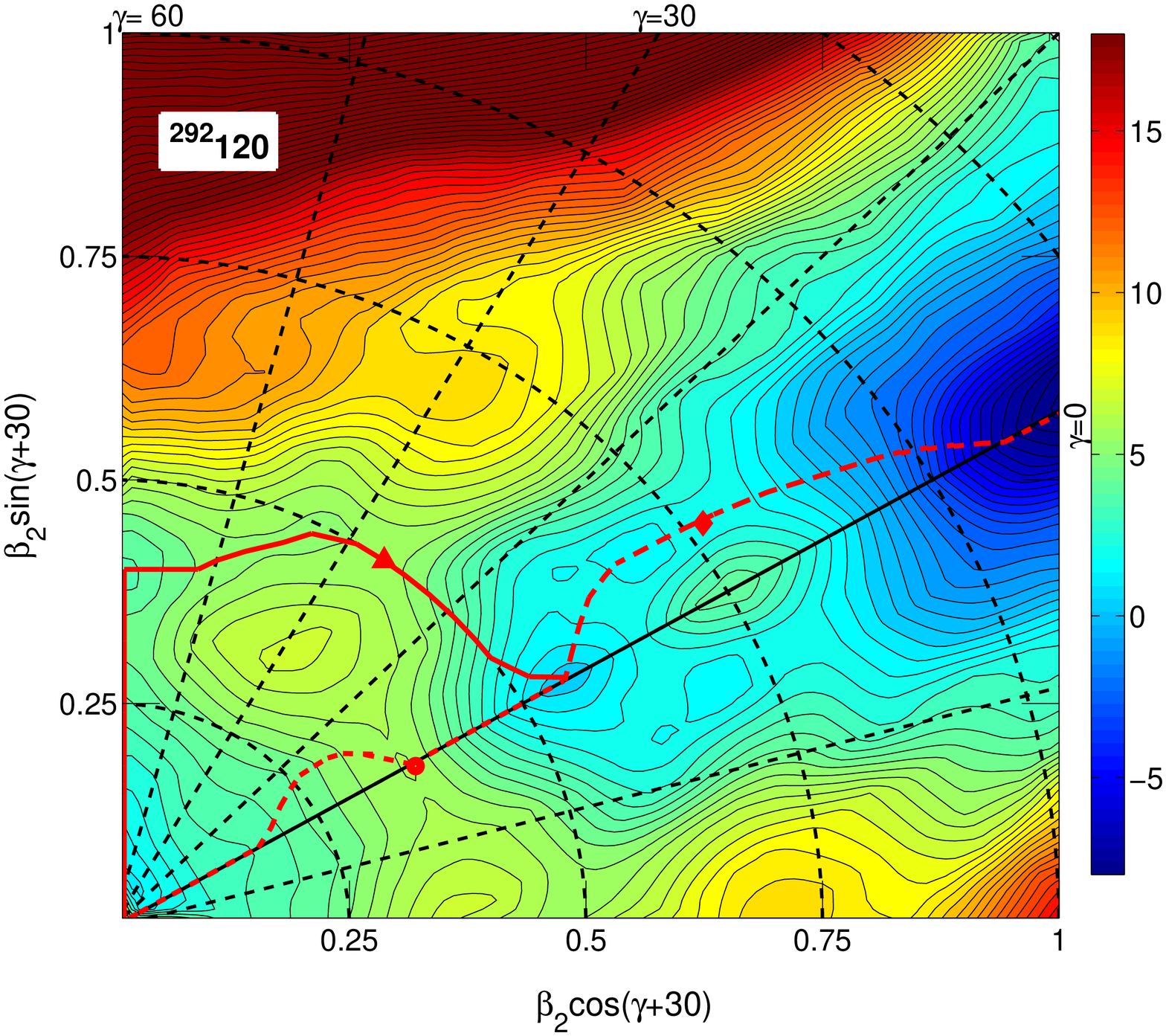}
\hspace{-0.2cm}
\includegraphics[width=6.3cm,angle=0]{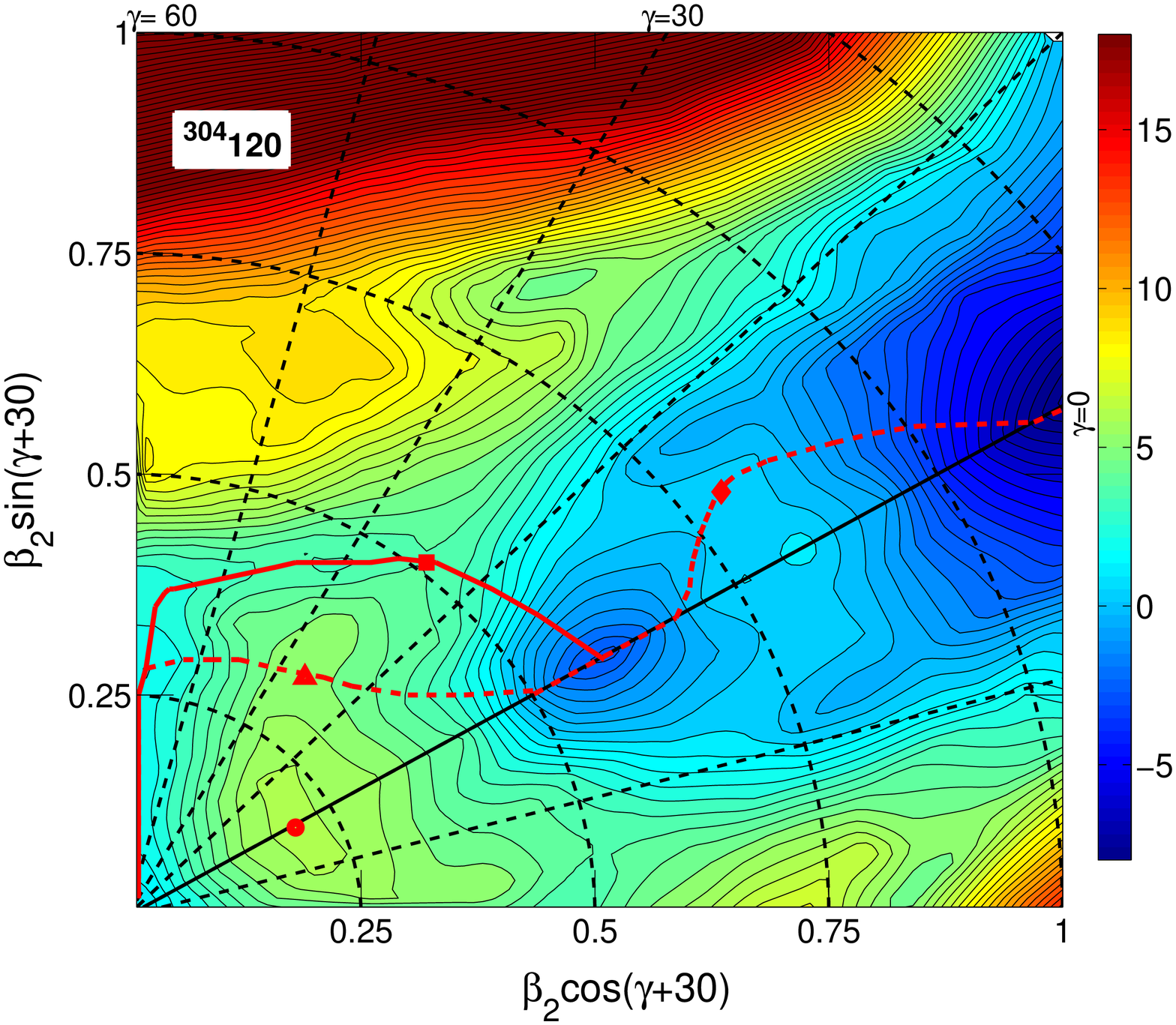}
\end{center}
\vspace{-0.3cm}
\caption{Potential energy surfaces of the $Z=120, N=172$ (left
panel) and $Z=120, N=184$ (right panel) nuclei. The energy 
difference between two neighboring equipotential lines is equal to 
0.5 MeV. The saddles are shown by solid symbols. The saddles are
defined via the immersion method (Ref.\ \protect\cite{MSI.09}), while the
fission path as a minimum energy path which represents the most probable 
pathway connecting two minima via a given saddle (see, for example, Ref.\ 
\protect\cite{STH.08}).}
\label{pes-2D}
\end{figure*}

  The situation is illustrated in Fig.\ \ref{pes-2D}. There are two fission pathes in the
doubly magic nucleus $^{292}$120.   The first fission path shown by a red dashed
line starts at spherical shape and then proceeds between two triaxial hills
located at moderate deformations
($\beta_2 \sim 0.35,\,\,\gamma \sim \pm30^{\circ}$) and bypasses the
axial hill at $\beta_2\sim$0.75 via a $\gamma \sim 7^{\circ}$ path.
The $\gamma$-softness of the PES, which exists between the two triaxial
hills, has only a minor effect on the shoulder of the inner fission barrier;
however, the height of inner fission barrier is not affected by triaxiality.
Second fission path shown by a solid red line starts at spherical shape,
and proceeds along the axially symmetric $\gamma=60^{\circ}$ axis,
via a saddle point at ($\beta_2 \sim 0.3$, $\gamma \sim 27^{\circ}$)
and then along the first fission path after second minimum. The unusual
physical feature of this second path is the fact that initially the nucleus
has to be squeezed along the axis of symmetry, thus creating an oblate
deformation with $\beta_2 \sim 0.35$. This is contrary to the usual picture
of fission were the nucleus is stretched out along the axis of
symmetry having a prolate deformation. A similar fission path also 
exists in the $^{304}$120 nucleus. However, the first fission path in this 
nucleus is modified due to the emergence of a axial hill at $\beta_2 \sim 0.2$; 
this shifts the saddle point from $(\beta_2 \sim 0.35, \gamma= 0^{\circ})$ in 
the $^{292}$120 nucleus to $(\beta_2 \sim 0.30, \gamma \sim 27^{\circ})$ in 
the $^{304}$120 nucleus.

  It is clear that the landscape of the PES of these nuclei in the region 
of the first fission barrier is more complicated than in the case of the actinides.
This calls for finding the dynamical path along which the fission process takes place.
Note that non-relativistic calculations based on the MM approach\cite{GSPS.99} showed that although triaxiality 
lowers the static fission barriers, it plays a minor role in spontaneous fission of 
superheavy nuclei with $Z\leq 120$. This is because a fission path via an oblate shape and 
triaxial saddles is substantially longer as compared with axially  
symmetric path which  leads to significant reduction of penetration probability.

  Note that contrary to actinide nuclei, the triaxiality has a
considerable impact on the shapes and the heights of outer fission
barriers of superheavy nuclei; the later are lowered by $\sim$ 2-3 MeV.

\vspace{-0.5cm}
\section{Conclusions}

   In conclusion, the level of accuracy of the description of
fission barriers in actinides is comparable for the methods
discussed here, the macroscopic+microscopic method, the
covariant and Gogny density functional
theories. Similar trends for the deviations from experiment as a function
of particle number seen in these approaches may indicate that further
substantial improvement can be achieved only in models which go beyond
mean field. The impact of the accuracy of the description of the
single-particle spectra in different models on the accuracy of reproduction
of fission barriers has also been discussed. It was shown that the landscape
of the potential energy surface in some superheavy nuclei is more
complicated than in the case of actinides. This indicates the need
for more investigations to determine the dynamical fission path in 
these nuclei.


This work has been supported by the U.S. Department of Energy under
the grant DE-FG02-07ER41459 and by the DFG cluster of excellence
\textquotedblleft Origin and Structure of the Universe
\textquotedblright\ (www.universe-cluster.de).



\begin{thebibliography}{99}

\bibitem{SP.07} A.\ Sobiczewski and K.\ Pomorski,
                {\it Prog.\ Part.\ Nucl.\ Phys.} {\bf 58}, 292  (2007).

\bibitem{AT.99} M.\ Arnould and K.\ Takahashi, {\it Rep.\ Prog.\ Phys.}
                {\bf 62},  395  (1999).

\bibitem{MPR.01} A.\ Mamdouh, J.\ M.\ Pearson, M.\ Rayet, and \
                 F.\ Tondeur, {\it Nucl.\ Phys.} {\bf A679}, 337  (2001).

\bibitem{DPS.04} J.\ Dudek, K.\ Pomorski, N.\ Schunck, and N.\ Dubray,
                 {\it Eur.\ Phys.\ J.} {\bf A20},  15 (2004).

\bibitem{AA.08} A.\ V.\ Afanasjev and H.\ Abusara, {\it Phys.\ Rev. C} {\bf 78},
                014315 (2008).

\bibitem{IOZ.02} M.\ G.\ Itkis, Y.\ T.\ Oganessian, and V.\ I.\ Zagrebaev,
                 {\it Phys.\ Rev.\ C} {\bf 65}, 044602  (2002).

\bibitem{VALR.05} D. Vretenar, A.~V. Afanasjev, G.~A. Lalazissis,
                  and P. Ring, {\it Phys. Rep.} {\bf 409},  101  (2005).

\bibitem{AA.10}  A.~V. Afanasjev and H. Abusara, {\it Phys. Rev. C} {\bf 81},
                 014309  (2010).

\bibitem{Gin.97} J.~N. Ginocchio, {\it Phys. Rev. Lett.} {\bf 78},  436
                 (1997).

\bibitem{AAR.10} H.\ Abusara, A.\ V.\ Afanasjev, and P.\ Ring,
                 {\it Phys. Rev. C}  {\bf 82} (2010)  044303.

\bibitem{BBM.04} T. B{\"u}rvenich, M. Bender, J.~A. Maruhn, and
                 P.-G. Reinhard, {\it Phys. Rev. C} {\bf 69},  014307 (2004).

\bibitem{KALR.10} S. Karatzikos, A.~V. Afanasjev, G.~A. Lalazissis,
               and P. Ring, {\it Phys. Lett.} {\bf B689},  72  (2010).

\bibitem{ARK.00} A.~V. Afanasjev, P. Ring, and J. K{\"o}nig, {\it Nucl.
                 Phys.} {\bf A676},  196 (2000).

\bibitem{NL3*} G.~A. Lalazissis, S. Karatzikos, R. Fossion,
               D. Pe{\~n}a~Arteaga, A.~V. Afanasjev, and
               P. Ring, {\it Phys. Lett.} {\bf B671},  36  (2009).

\bibitem{DPB.07} J.\ Dobrowolski, K.\ Pomorski, and J.\ Bartel,
                 {\it Phys.\ Rev. C} {\bf 75},  024613 (2007).

\bibitem{MSI.09} P.\ M{\"o}ller, A.\ J.\ Sierk, T.\ Ichikawa,
                 A.\ Iwamoto, R.\ Bengtsson, H.\ Uhrenholt, and
                 S.\ {\AA}berg, {\it Phys.\ Rev. C} {\bf 79}, 064304 (2009).

\bibitem{DGGL.06} J.-P.\ Delaroche, M.\ Girod, H.\ Goutte, and
                  J.\ Libert, {\it Nucl.\ Phys.} {\bf A771}, 103 (2006).

\bibitem{RIPL-2} RIPL-2 stands for reference input parameter library of
International Atomic Energy Agency located at http://www-nds.iaea.org/ripl2/,
which for actinides is based on Ref.\protect\cite{M.98}.

\bibitem{M.98} V.\ M.\ Maslov. {\it RIPL-1 Handbook.} TEXDOC-000, IAEA,
               Vienna, 1998, Ch.\ 5.

\bibitem{MM.11} D.\ G.\ Madland and P.\ M{\"o}ller, {\it Los Alamos National
Laboratory unclassified report}, LA-UR-11-11447 (2011).

\bibitem{BHBG.74} B.\ B.\ Back, O.\ Hansen, H.\ C.\ Britt, and
                  J.\ D.\ Garrett,  {\it Phys.\ Rev. C} {\bf 9}, 1924 (1974).

\bibitem{B.80} H.\ C.\ Britt, in {\it Proc. of the Symposium on the
Physics and Chemistry of Fission,} J{\"u}lich, Germany, 14-18 May 1979
(IAEA, Vienna, 1980), Vol.\ I, p.\ 3.

\bibitem{BL.80} S.\ Bjornholm and J.\ E.\ Lynn, {\it Rev.\ Mod.\ Phys.} 
                {\bf 52}, 725 (1980) and references therein.

\bibitem{A250}  A.\ V.\ Afanasjev {\it et al},
{\it Phys.\ Rev. C}
                {\bf 67}, 024309 (2003).

\bibitem{SDMMNSS.10} N.\ Schunck, J.\ Dobaczewski, J.\ McDonnell,
               J.\ Mor{\'e}, W.\ Nazarewicz, J.\ Sarich, and M.\ V.\ Stoitsov,
               {\it Phys.\ Rev. C} {\bf 81}, 024316 (2010).

\bibitem{RSR.10} R.\ Rodriguez-Guzman, P.\ Sarriguren, and L.\ M.\ Robledo,
                 {\it Phys.\ Rev. C} {\bf 82}, 061302(R) (2010).

\bibitem{AS.11} A.\ V.\ Afanasjev and S.\ Shawaqfeh, {\it Phys.\ Lett.} 
                {\bf B706}, 177 (2011).

\bibitem{Sol-book2} V.\ G.\ Soloviev, {\it Theory of Atomic Nuclei:
                    Quasiparticles and Phonons} (Institute of Physics
                    Publishing, Bristol and Philadelphia, 1992).

\bibitem{RW.73} P.\ Ring and E.\ Werner, {\it Nucl. Phys.} {\bf A211},
                198 (1973).

\bibitem{LR.06} E.\ Litvinova and P.\ Ring, {\it Phys.\ Rev. C} {\bf 73},
                044328 (2006).

\bibitem{LA.11} E.\ V.\ Litvinova and A.\ V.\ Afanasjev,
                {\it Phys.\ Rev. C} {\bf 84}, 014305 (2011).

\bibitem{JSSJ.90} A.\ K.\ Jain, R.\ K.\ Sheline, P.\ C.\ Sood, and
                  K.\ Jain, {\it Rev.\ Mod.\ Phys.} {\bf 62}, 393 (1990).

\bibitem{PS.04} A.\ Parkhomenko and A.\ Sobieczewski, {\it Acta Phys.\
                Polonica} {\bf 35}, 2447 (2004).

\bibitem{AAR.11} H.\ Abusara, A.\ V.\ Afanasjev, and P.\ Ring,
                 submitted to Phys. Rev. C.

\bibitem{GSPS.99} R.\ A.\ Gherghescu, J.\ Skalski, Z.\ Patyk, and
                  A.\ Sobiczewski, {\it Nucl.\ Phys.} {\bf A651} (1999) 237.

\bibitem{STH.08} D.\ Sheppard, R.\ Tyrell and G.\ Henkelman, {\it Jour.\ Chem.\ Phys.} {\bf 128} (2008) 134106. 

\end{thebibliography}
\end{document}